\documentclass{svproc}
\usepackage{multirow}
 \usepackage{graphicx}
 \usepackage{amsfonts}
 \usepackage{amsmath}
 \usepackage{amssymb}

\usepackage{url}

\begin{document}
\mainmatter            
\title{Music Therapy based Stress Prediction using Homological Feature Analysis on EEG Signals}
\titlerunning{Music Therapy} 
\author{Srikireddy Dhanunjay Reddy, Tharun Kumar Reddy Bollu}
\authorrunning{SD Reddy et al.} 
\institute{Department of Electronics \& Communication Engineering\\Indian Institute of Technology, Roorkee.\\
\email{\{sd\_reddy,tharun.reddy\}@ece.iitr.ac.in}
}

\maketitle          
\begin{abstract}
Stress became a common factor in the busy daily routines of all academic and corporate working environments. Everyone checks for efficient stress-buster alternatives to calm down from work pressure. Instead of investing time in unnecessary efforts, this work shows the stress relief scenario of subjects by listening to "Raag Darbari" music notes as a simple add-on to their schedule. An innovative approach has been implemented on the "MUSEI-EEG" dataset using Topological Data Analysis (TDA) to analyze this stress relief study. This study reveals that persistent homological features can be robust biomarkers for classifying closely distributed subject data. The proposed TDA approach framework revealed homological features like birth-death rate and entropy efficacy in stress prediction using Electroencephalogram (EEG) signals with 86\% average accuracy and 0.2 standard deviation.
\end{abstract}
\section{Introduction}
EEG signal is one of the non-invasive neuro-imaging modalities, to visualize the electrical activity of neurons in the brain. Understanding the behavior of these signals can be useful for monitoring and diagnosis of the mental health condition \cite{i1}. By Gallup Global Reports, 2021, the World Health Organization (WHO) revealed that 190 million people around the world are feeling stressed based on their work or academic environments \cite{i2}. This made the urge to increase stress relief-based studies for individuals' mental health and well-being. Related to this, stress rehabilitation or relief assessment studies have been done by utilizing the EEG signal behavior \cite{i3}. Many methods are being followed as add-ons of daily routines to reduce stress, among them, stress relief through dance, virtual reality interaction, meditation, yoga, etc., gave fruitful results \cite{i4}-\cite{i7}. \\ \hspace{0pt}
A wide range of physiological signal analyses have been done for anomaly detection, including wavelet, multivariate, and Riemannian-based techniques \cite{i8}-\cite{i10}. Recent studies find that TDA techniques can be useful biomarker tools for EEG signal processing \cite{i11}. Persistent homological features like birth-death rates, barcodes, betti curves, and landscapes of groups formed by the point-cloud simplexes reveal the internal behavior of the multi-dimensional data series \cite{i12}. Homological features extracted from these techniques are robust biomarkers for cognitive state characterization \cite{i13}. In addition to these, simplicial complex generation techniques like Vietoris-Rips complex generation and Voronoi-Tessellation formation give prominent results for mental health disorder monitoring \cite{i14} \cite{i15}. Extracted persistent homological features can be given to the machine learning (ML) models for the computer-aided diagnosis application. \\ \hspace{0pt}          
Key contributions for stress prediction through this paper are as follows:
\begin{itemize}
    \item To the best of our knowledge, a unique 9-minute three-stage music paradigm is designed for the analysis and assessment of stress in individuals.
    \item TDA-based framework is implemented to know the homological feature ability for classifying closely distributed class data.
\end{itemize}
The rest of this paper's organization follows this flow, section 2 describes the MUSEI-EEG dataset used for the stress prediction. Section 3 explains the proposed methodology with mathematical validation, then section 4 discusses the simulated results and ML model performances of the proposed methodology. Finally, section 5 concludes the paper with noticed findings.
\section{Dataset Description}
The proposed TDA framework is implemented on the MUSEI-EEG dataset \cite{d1}, which comprises 20 undergraduate subjects' EEG data to identify stress relief after hearing Raag Darbari music. 8 female and 12 male subjects aged 18 to 21 years, participated in the designed 9-minute three-stage music paradigm. There exists pre-task, task, and post-task segments in the whole paradigm. Each segment duration is about 3 minutes, subjects are guided to keep a calm and relaxed eyes-closed state in both pre and post-task segments. But, in the task segment stage, subjects are set to listen to a 3-minute Raag Darbari track in relax mode without any movement. EEG signals have been recorded all over three segments separately using the Brain Vision actiCHamp system (32 channels) at 500 Hz sampling frequency. A standard 10-20 system is used for the electrode placement without noise rejection scenario and time-stamps.\\
In addition to physiological signal monitoring, participants filled out the stress anxiety scale (Y1), trait anxiety scale (Y2), and emotional intelligence (EQ) forms, both prior to and after the execution of the designed paradigm. Based on the averaged scale indexes of these three forms, baseline labels are assigned and further frameworks can be done for the stress prediction using the EEG signals.
\section{Proposed Methodology}
\subsection{Preprocessing}
\begin{figure}[h!]
    \centering
    \includegraphics[width=\textwidth]{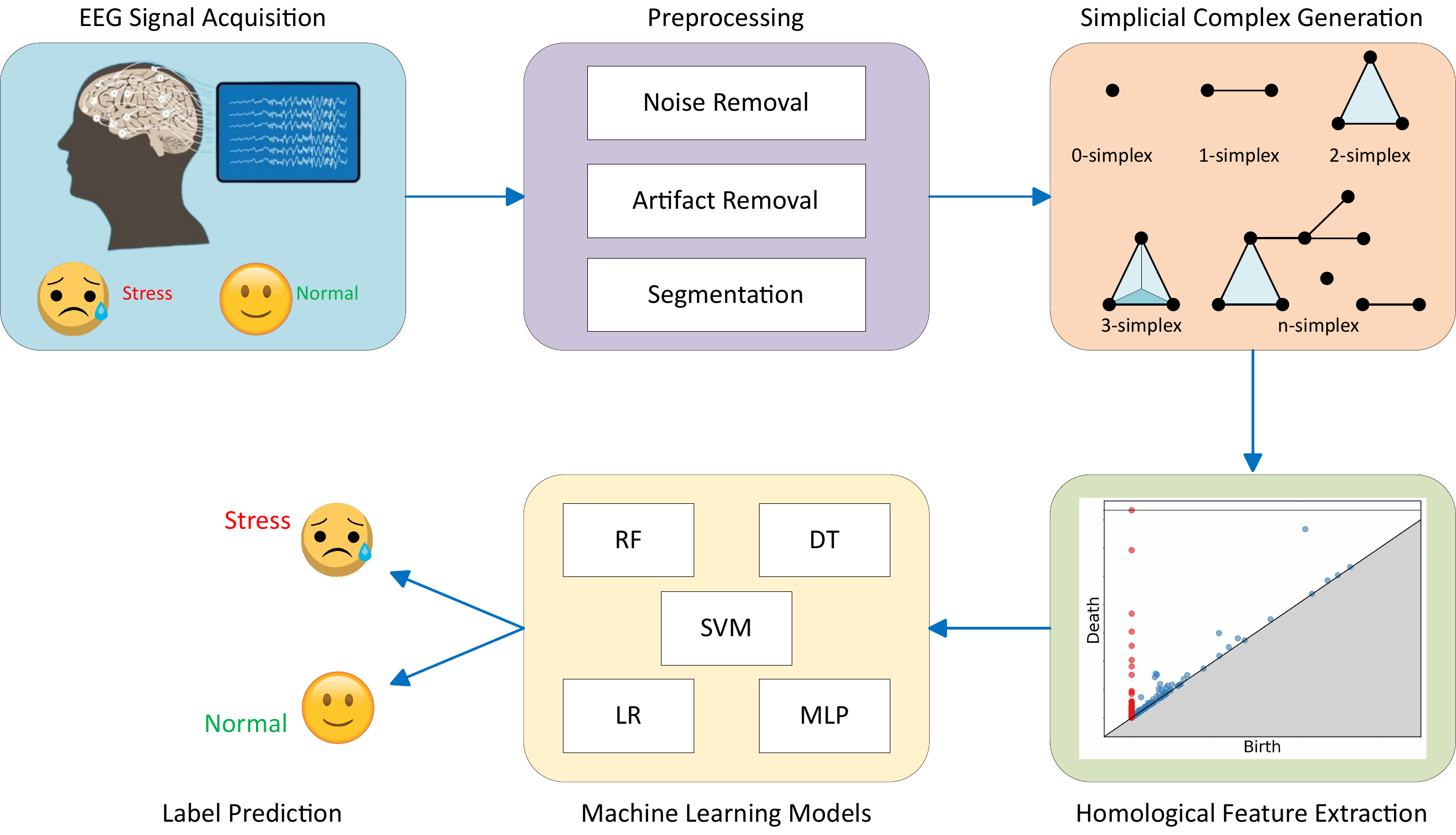} 
    \caption{Proposed framework for music therapy based stress prediction}
    \label{framework} 
\end{figure}

A TDA approach framework is implemented on the MUSEI-EEG dataset for the homological feature-based analysis as shown in Figure \ref{framework}. Further, in this section, detailed implementation is discussed with the relevant mathematical validation. 
Consider the EEG signal recording of a subject, $\mathcal{X} \in \mathbb{R}^{c \times n}$ with, $c$ number of channels and $n$ number of time samples. Further, $\mathcal{X}$ is subjected to the preprocessing stage to remove the additional noise, and artifacts added during the signal acquisition. Noise removal has been done using the bandpass filter ($\mathcal{B}$) with the frequency range [0.5 ($f_l$)-60 ($f_h$)] Hz, in addition to this, $\mathcal{X}$ is passed through a notch filter ($\mathcal{N}$) to remove the power line interference (PLI) at 50 Hz, to get final filtered signal $\mathcal{X}^\prime$.
\begin{equation}
    \mathcal{X}^\prime = \mathcal{N}(\mathcal{B}[\mathcal{X},f_l,f_h],50)
\end{equation}
Now, the preprocessed EEG $\mathcal{X}^\prime$ is segmented into $k$ 10-second trials. The final segmented EEG signal $\tilde{\mathcal{X}}\in \mathbb{R}^{k \times c \times t}$ with $t$ samples in each trial is processed further to extract the persistent homological features.
\subsection{Simplicial Complex Generation}
\begin{figure}[h!]
    \centering
    \includegraphics[width=\textwidth]{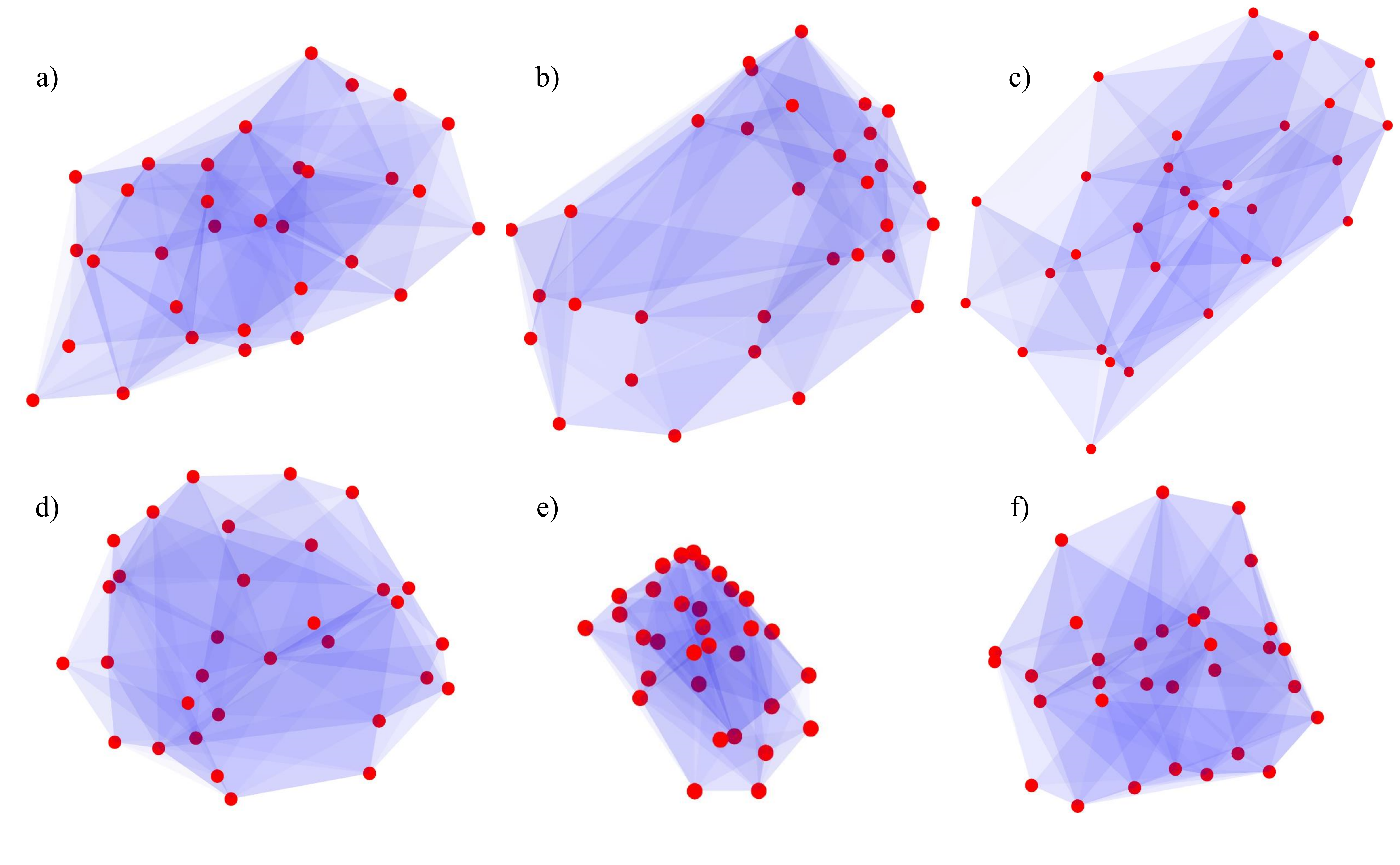} 
    \caption{Generated simplicial complexes of, Up: stress and Down: normal subjects. a) \& d) for  pre-task, b) \& e) for task, and c) \& f) for post-task segments}
    \label{complex} 
\end{figure}

Principal component analysis (PCA) and multi-dimensional scaling (MDS) can be efficient techniques for the point cloud $\mathcal{P}_c \in \mathbb{R}^{c \times d}$ in the topological metric space $\mathbb{R}^d$ \cite{m1} \cite{m2}. Based on the application and the computational power, $d$ value can be selected. In this work, three-dimensional metric space has been generated using the PCA technique to analyze the distance-based distribution of the data points in $\mathcal{P}_c$. Any trial from the $\tilde{\mathcal{X}}$ can be decomposed into its eigen and orthonormal components as shown in equation (2). $\mathcal{P}_c$ is generated by mapping the d-eigenvalues $\Lambda_d$ onto the orthonormal vectors $\mathcal{V}_d$.    
\begin{equation}
    \tilde{\mathcal{X}} = \mathcal{V} \Lambda \mathcal{V}^T
\end{equation}
\begin{equation}
    \mathcal{P}_c= \Lambda_d \mathcal{V}_d
\end{equation}
Obtained, $\mathcal{P}_c:\{p_1, p_2,\dots, p_c\}$ consists of $c$ number of data points, each representing the individual EEG channel. This leads to an analysis of connectivity analysis of channels using homological features. Moreover, a simplicial complex $\mathcal{S}$ needs to be generated for the extraction of the persistent homological features from Homological groups $(H_0, H_1, H_2)$ \cite{m2}. To generate $\mathcal{S}$, consider the ball/ disc ($b_r(.)$) of radius ($r$) around any data point $p_i$. $r$ of $b_r(p_i)$ is increased until it reaches the threshold value. The threshold value of the ball radius is defined based on the minimum distance between the data points ($\epsilon$) present in $\mathcal{P}_c$ as shown in equation (4).
\begin{equation}
    b_r(p_i) \leq 2 \epsilon; \epsilon= min(\delta(p_i,p_j))
\end{equation}
$p_i$ and $p_j$ can be any data points of $\mathcal{P}_c$, and $\delta(p_i,p_j)$ is the euclidean distance between any two points, i.e. $||p_i-p_j||^2$. In this increasing radius process, the evolution of the homological groups can be saved as the persistent homological features. In the final, generated $\mathcal{S}$ can be represented as shown in equation (5). Figure \ref{complex}, displays the generated $\mathcal{S}$ for stress and normal subjects in all three segments of the data acquisition. 
\begin{equation}
   \mathcal{S}\rightarrow b_r(p_i) \bigcap b_r(p_j) \neq \phi\Leftrightarrow\delta(p_i,p_r)\leq 2\epsilon 
\end{equation}
\subsection{Homological Feature Extraction}
Persistent homological features used in this work are the birth and death rates $(x,y)$ of ($H_0, H_1, H_2$), during the process of $\mathcal{S}$ generation. Simplexes shown in the Figure \ref{framework} are the building blocks of ($H_0, H_1, H_2$). The mathematical expression of these features can be given as equation (6).
\begin{equation}
    f(x, y) = \{(x, y) : x, y \in \mathbb{R}\} \cup \{(x, \infty) : x \in \mathbb{R}\}
\end{equation}
Furthermore, to give more information about each subject to ML models, persistent entropy $(\mathcal{E})$ is also extracted from the $f(x,y)$. 
\begin{equation}
    \mathcal{E} = \sum_{l=1}^{L} H \log(H); \quad H =\frac{x_l-y_l} {\sum_{l=1}^{L} (x_l - y_l)}
\end{equation}
$L$ represents the number of homological groups that exist in each $\mathcal{S}$. Discussed features are extracted for all the 20 subjects by following the framework as shown in Figure \ref{framework}. In the final, extracted features are given to the ML models, and the best-performed model performance is compared in Table \ref{model comparison}. Extracted features are visually inspected to analyze the behavior of stress and normal subjects, and the discussion of $\mathcal{S}$ and $(x,y)$ distribution and its relevance to stress detection is discussed in further sections. 
\section{Results \& Discussion}
Homological features are extracted from the EEG signals of the MUSEI-EEG dataset using the TDA framework discussed in the methodology section. After preprocessing the collected signals, $\mathcal{S}$ is generated as shown in Figure \ref{complex}, it represents the $\mathcal{S}$ of stress (up) and normal (down) subjects. In particular, these subjects are in the same state through all three segments of the designed music paradigm. Visually, it is observable that the distance between the data points of $\mathcal{S}$ is greater in the stress subject than in the normal subject. With this observation, visually, stress prediction can be made based on the arrangement of data points (EEG channels) in the topological point cloud space and its simplicial complex. Further, extracted features can be given to ML models for the subject label prediction.\\ \hspace{0pt}
\begin{figure}[h!]
    \centering
    \includegraphics[width=\textwidth]{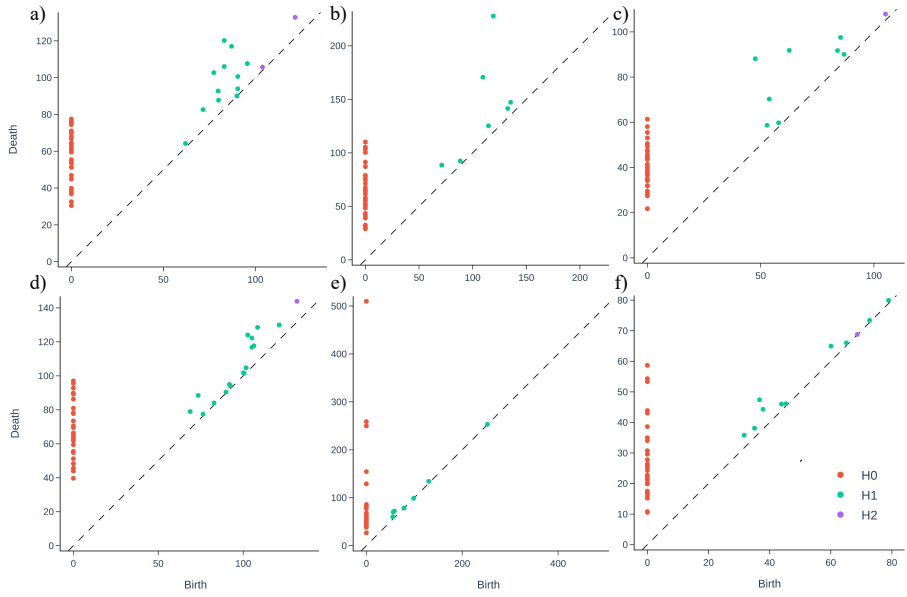} 
    \caption{Persistent homological feature plots of, Up: stress and Down: normal subjects. a) \& d) for  pre-task, b) \& e) for task, and c) \& f) for post-task segments}
    \label{homology features} 
\end{figure}
Figure \ref{homology features} shows the birth-death rates $(x,y)$ of $(H_0, H_1, H_2)$, in the evolution process of $\mathcal{S}$ using the proposed TDA framework. In relevance with the $\mathcal{S}$ shown in Figure \ref{complex}, homological features $(x,y)$ are plotted for the same subjects. Homological groups settling towards the diagonal line or $\infty$-line indicate the stability of the data point distribution. Figure \ref{homology features} demonstrates the stable distribution of normal subject homology features in all three segments, but not in stress subject features. With this feature distribution's nature, homological features can be utilized as biomarkers for stress prediction. \\ \hspace{0pt}   
\begin{table}[ht]
\centering
\caption{Comparison model performances on stress classification of the MUSEI-EEG dataset }
\label{model comparison}
\resizebox{\columnwidth}{!}{%
\begin{tabular}{|c|ccccccccc|}
\hline
\multirow{3}{*}{\textbf{Model}} & \multicolumn{9}{c|}{\textbf{MUSEI-EEG   segments}}                                                                                                                                                                                                                       \\ \cline{2-10} 
                                & \multicolumn{3}{c|}{\textbf{Pre-task}}                                                        & \multicolumn{3}{c|}{\textbf{Task}}                                                            & \multicolumn{3}{c|}{\textbf{Post-task}}                                  \\ \cline{2-10} 
                                & \multicolumn{1}{c|}{Accuracy±SD} & \multicolumn{1}{c|}{f1 score} & \multicolumn{1}{c|}{Kappa} & \multicolumn{1}{c|}{Accuracy±SD} & \multicolumn{1}{c|}{f1 score} & \multicolumn{1}{c|}{Kappa} & \multicolumn{1}{c|}{Accuracy±SD} & \multicolumn{1}{c|}{f1 score} & Kappa \\ \hline
\textbf{RF}                     & \multicolumn{1}{c|}{0.89±0.012}  & \multicolumn{1}{c|}{0.86}     & \multicolumn{1}{c|}{0.8}   & \multicolumn{1}{c|}{0.88±0.014}  & \multicolumn{1}{c|}{0.85}     & \multicolumn{1}{c|}{0.79}  & \multicolumn{1}{c|}{0.82±0.015}  & \multicolumn{1}{c|}{0.84}     & 0.77  \\ \hline
\textbf{DT}                     & \multicolumn{1}{c|}{0.86±0.0268} & \multicolumn{1}{c|}{0.85}     & \multicolumn{1}{c|}{0.91}  & \multicolumn{1}{c|}{0.79±0.021}  & \multicolumn{1}{c|}{0.8}      & \multicolumn{1}{c|}{0.88}  & \multicolumn{1}{c|}{0.80±0.019}  & \multicolumn{1}{c|}{0.77}     & 0.85  \\ \hline
\textbf{SVM}                    & \multicolumn{1}{c|}{0.76±0.0156} & \multicolumn{1}{c|}{0.82}     & \multicolumn{1}{c|}{0.84}  & \multicolumn{1}{c|}{0.75±0.013}  & \multicolumn{1}{c|}{0.8}      & \multicolumn{1}{c|}{0.81}  & \multicolumn{1}{c|}{0.72±0.020}  & \multicolumn{1}{c|}{0.78}     & 0.79  \\ \hline
\textbf{LR}                     & \multicolumn{1}{c|}{0.88±0.026}  & \multicolumn{1}{c|}{0.92}     & \multicolumn{1}{c|}{0.94}  & \multicolumn{1}{c|}{0.86±0.018}  & \multicolumn{1}{c|}{0.89}     & \multicolumn{1}{c|}{0.9}   & \multicolumn{1}{c|}{0.78±0.016}  & \multicolumn{1}{c|}{0.86}     & 0.87  \\ \hline
\textbf{MLP}                    & \multicolumn{1}{c|}{0.84±0.012}  & \multicolumn{1}{c|}{0.86}     & \multicolumn{1}{c|}{0.82}  & \multicolumn{1}{c|}{0.81±0.023}  & \multicolumn{1}{c|}{0.83}     & \multicolumn{1}{c|}{0.8}   & \multicolumn{1}{c|}{0.79±0.020}  & \multicolumn{1}{c|}{0.81}     & 0.79  \\ \hline
\end{tabular}%
}
\end{table}
The extracted homological features $(x,y, \mathcal{E})$ are given to different ML models for label prediction purposes. Among them, Random Forest (RF), Decision Tree (DT), Support Vector Machine (SVM), Logistic Regression (LR), and Multi-Layer Perceptron performed well with a 5-fold cross-validation scheme. 80-20 train-test split and parameter tuning are implemented on these models, performance metric comparison is tabulated in Table \ref{model comparison}. Moreover, a statistical significance p-test is also performed on Y1, Y2, and EQ forms for ground truth label validation. As a result, it shows the significance as $p < 0.01$ for Y1 Vs EQ and Y2 Vs EQ, $p<0.5$ for Y1 Vs Y2.\\ \hspace{0pt}   
\begin{figure}[h!]
    \centering
    \includegraphics[width=\textwidth]{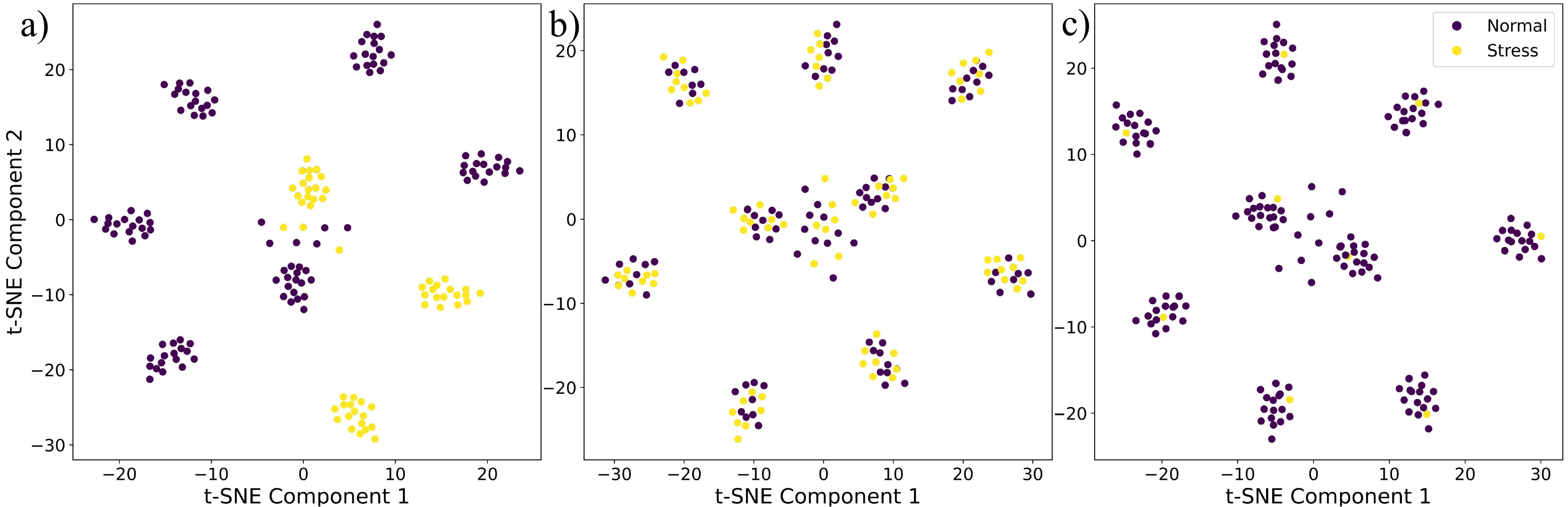} 
    \caption{Generated t-SNE embeddings of stress and normal subjects at a) pre-task, b) task, and c) post-task segments}
    \label{embeddings} 
\end{figure}
Finally, to understand the EEG signal distributional behavior, t-SNE components are taken out and plotted as shown in Figure \ref{embeddings} for all three segments of the music paradigm. It can be observed that in the case of task and post-task segments, stress and normal subjects are closely distributed. This scenario may be lead the ML model into ambiguity in stress prediction. Performance metrics comparison in Table \ref{model comparison} shows the stable ML model performances in both task and post-task segments, as same as the pre-task segment. From this observation and the performance, homological features can be robust biomarkers for the prediction application in closely distributed data sample cases. 
\section{Conclusion}
In this paper, a TDA approach framework is implemented on a unique 9-minute three-stage music paradigm for stress prediction among the subjects using EEG signals. The designed paradigm shows that Raag Darbari's music listening can be taken as a stress relief application to calm ourselves in academic and work environment pressures. Moreover, the proposed TDA framework shows the ability of the homological features as robust biomarkers for stress prediction among closely distributed nature subjects. In the future, advanced TDA techniques and other non-Euclidean techniques like Riemannian manifolds can be implemented to improve stress prediction analysis on similar kinds of EEG signal data.
\section*{Acknowledgement}
In this article, MUSEI-EEG data collection was approved by the Institute Human Ethics Committee (IHEC) of the Indian Institute of Technology, Roorkee, with permission grant IITR/IIC/22/2-12.

\end{document}